\documentclass[manuscript]{aastex}

\usepackage{mathbbol}

\shorttitle{Background solar wind modeling}
\shortauthors{Wiengarten et al.}

\begin{document}

\title{Cosmic Ray Transport in Heliospheric Magnetic Structures:\\ I. Modeling Background Solar Wind Using the CRONOS MHD Code}


\author{T. Wiengarten\altaffilmark{1}, J. Kleimann\altaffilmark{1} and H. Fichtner\altaffilmark{1}}
\affil{Institut f\"ur Theoretische Physik IV, Ruhr-Universit\"at Bochum, Germany}

\author{P. K\"uhl\altaffilmark{2}, A. Kopp\altaffilmark{2}, and B. Heber\altaffilmark{2}}
\affil{Institut f\"ur Experimentelle und Angewandte Physik, Christian-Albrecht-Universit\"at zu Kiel, Germany}

\and

\author{R. Kissmann\altaffilmark{3}}
\affil{Institut f\"ur Astro- und Teilchenphysik, Universit\"at Innsbruck, Austria}




\begin{abstract}
The transport of energetic particles such as Cosmic Rays is governed by the properties of the plasma being traversed. While these properties are rather poorly known for galactic and interstellar plasmas due to the lack of in situ measurements, the heliospheric plasma environment has been probed by spacecraft for decades and provides a unique opportunity for testing transport theories. Of particular interest for the 3D heliospheric transport of energetic particles are structures such as corotating interaction regions (CIRs), which, due to strongly enhanced magnetic field strengths, turbulence, and associated shocks, can act as diffusion barriers on the one hand, but also as accelerators of low energy CRs on the other hand as well. 
In a two-fold series of papers we investigate these effects by modeling inner-heliospheric solar wind conditions with a numerical magnetohydrodynamic (MHD) setup (this paper), which will serve as an input to a transport code employing a stochastic differential equation (SDE) approach (second paper).
In this first paper we present results from 3D MHD simulations with our code CRONOS: for validation purposes we use analytic boundary conditions and compare with similar work by Pizzo. For a more realistic modeling of solar wind conditions, boundary conditions derived from synoptic magnetograms via the Wang-Sheeley-Arge (WSA) model are utilized, where the potential field modeling is performed with a finite-difference approach (FDIPS) in contrast to the traditional spherical harmonics expansion often utilized in the WSA model. Our results are validated by comparing with multi-spacecraft data for ecliptical (STEREO-A/B) and out-of-ecliptic (Ulysses) regions.

\end{abstract}



\section{Introduction}
\subsection{Cosmic ray transport}
The understanding and appropriate modeling of the transport of charged energetic particles 
in turbulent magnetic fields remains one of the long-standing challenges in astrophysics and space 
physics. The many simultaneous in situ observations of both the heliospheric magnetic
field and different energetic particle populations make the heliosphere a natural laboratory
for corresponding studies. Of particular interest in this context are galactic cosmic rays (GCRs)
and Jovian cosmic ray electrons. While the former traverse the whole three-dimensionally structured
heliosphere and allow for studies of its large-scale variations as well as of the evolution of 
heliospheric turbulence \citep[e.g.,][]{Zank-etal-1996, Heber-etal-2006, Potgieter-2013a, Bruno-Carbone-2013}, the latter represent 
a point-like source and are, thus, well suited for analyses of anisotropic spatial diffusion 
\citep[e.g.,][]{Ferreira-etal-2001a,Ferreira-etal-2001b, Sternal-etal-2011, Strauss-etal-2013}.\\
In order to exploit this natural laboratory in full, it is necessary reproduce the measurements 
with simulations that do contain as much as possible of the three-dimensional structure of the
plasma background within which the cosmic rays are propagating. Significant progress has been
made regarding the modeling of the quiet solar wind \citep[e.g.,][]{Potgieter-2013b} but much remains to be done
to implement the many features that are structuring the solar wind and the heliospheric magnetic
field, into transport models of cosmic rays. Particularly interesting structures are the 
corotating interaction regions (CIRs) that are formed during the interaction of fast solar wind
streams from coronal holes with preceding slow solar wind and usually persist for several solar 
roations \citep[e.g.,][and references therein]{Balogh-etal-1999}. These structures not only lead to particle acceleration,
but also to the modulation of GCR and Jovian electron spectra \citep{Richardson-2004}. Indeed
Ulysses measurements, as described in, e.g., \citet{Marsden-2001}, at high heliolatitudes during the
first orbit of the spacecraft around the Sun indicated that CIRs represent the major constituent
for the three-dimensional heliospheric structure close to solar minimum 
\citep[e.g.][]{Heber-etal-1999b}. A major surprise came from the measurements of accelerated 
energetic particles and GCRs that showed clear periodic signals even above the poles of the Sun
where no in situ signals of CIRs were registered by the plasma or magnetic field instrument
\citep{Kunow-etal-1995}. Electron measurements, however, indicate no variation at these region.
It can be speculated that the differences are caused by the fact that both GCRs and locally 
accelerated particles have an extended source, while MeV electrons in the inner heliosphere 
originate from a point-like source as mentioned above \citep{Chenette-1980}. Thus, there is a
different influence of CIRs on GCR and Jovian electron flux variations as investigated recently by
\citet{Kuehl-etal-2013}.\\
If the electron source is a well-localized point-like source in the heliosphere --- namely the Jovian 
magnetosphere --- it is mandatory to describe the structure of the plasma stream in the whole inner
heliosphere up to several AU and not only at the location of different spacecraft measuring these
particles. 
\begin{figure}
\includegraphics[width=0.5\textwidth]{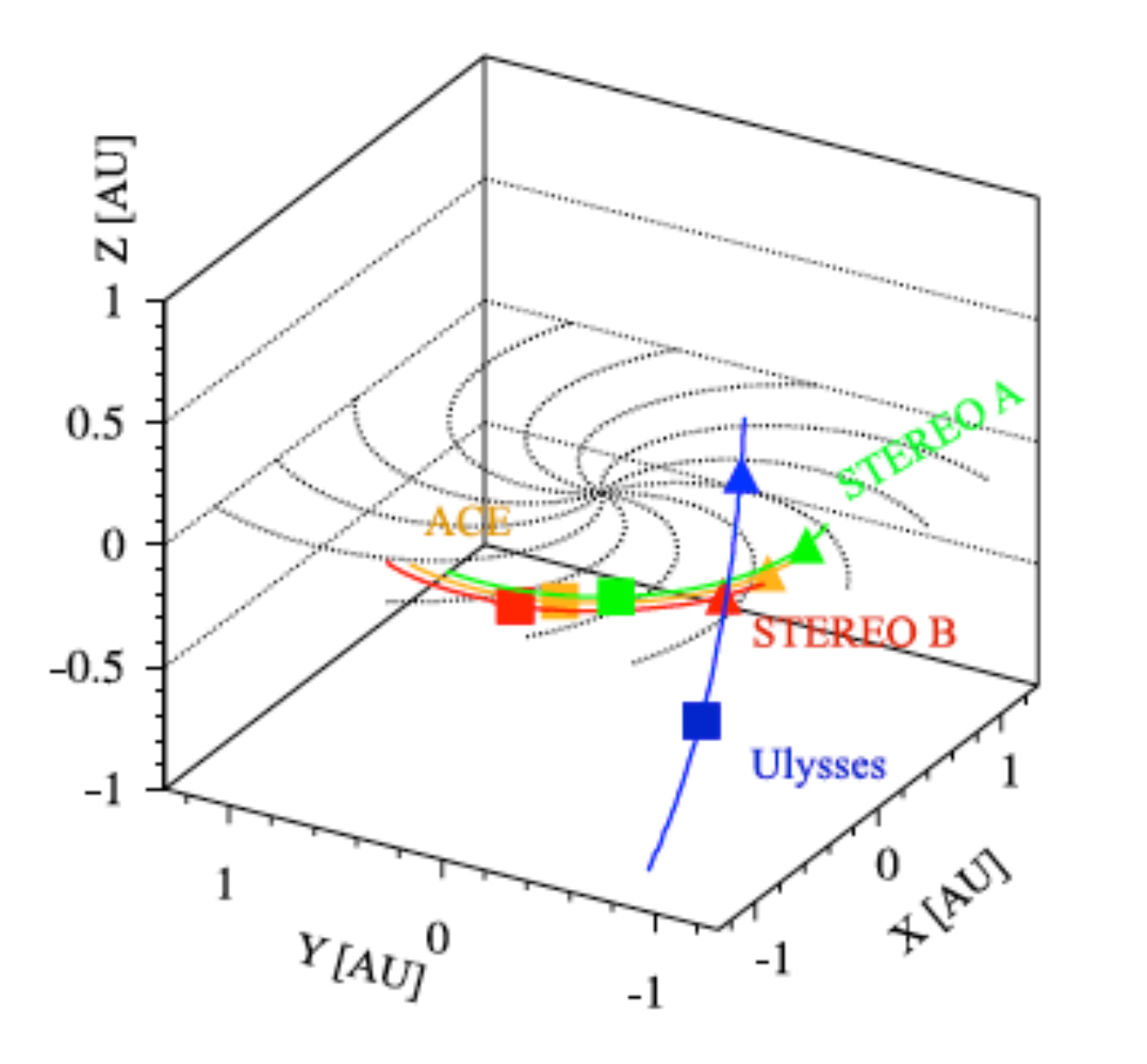}
\includegraphics[width=0.5\textwidth]{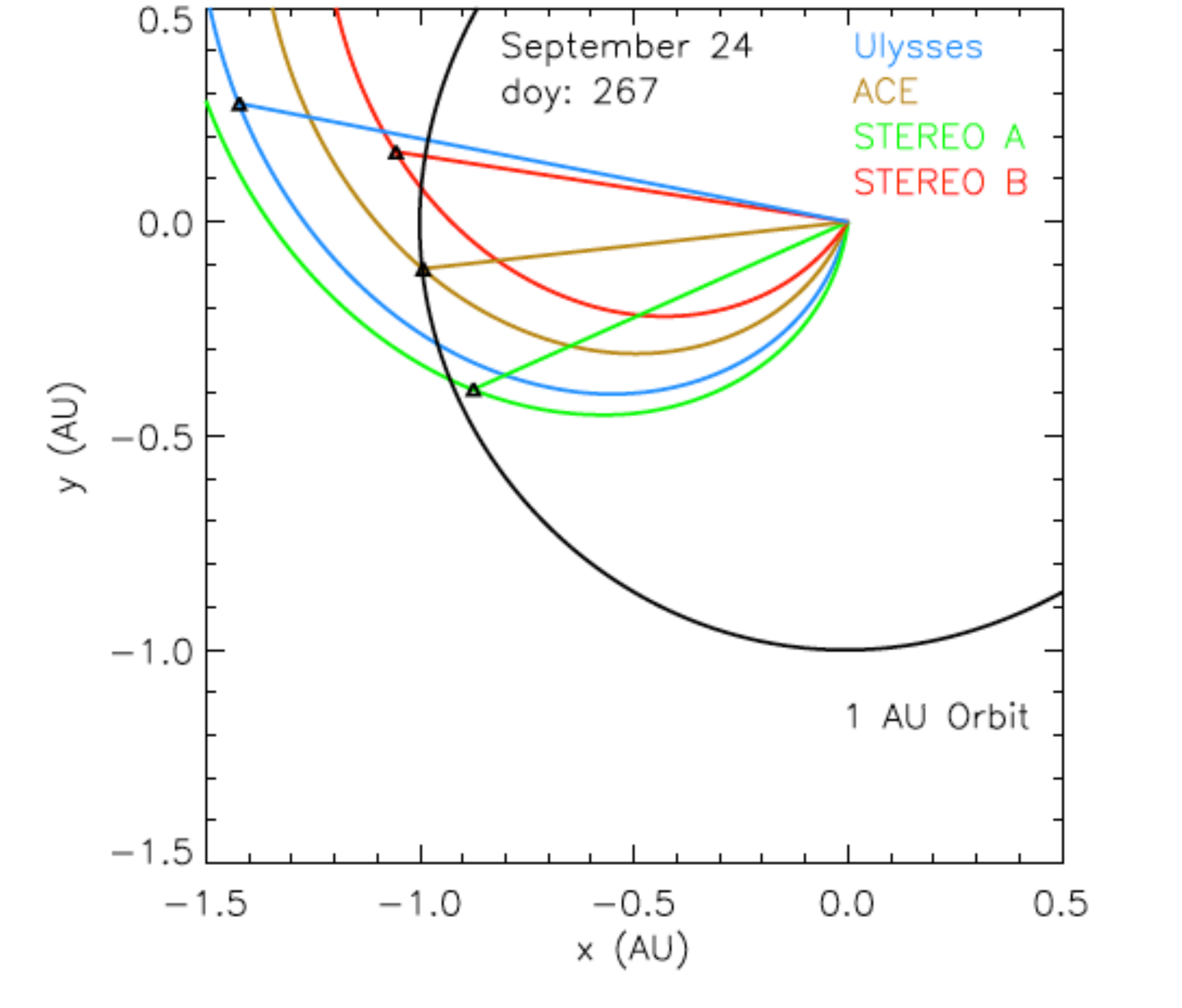}
\caption{Left: Ulysses (blue), STEREO-B (red), SOHO (gold) and STEREO-A (green) trajectories in ecliptic coordinates. The dotted curves display nominal Parker spirals for 400 km s$^{-1}$ separated by 30 degrees. Squares and triangles mark the positions of the spacecraft for doy 213 and 267, respectively. Right: Projection of trajectories in the ecliptic plane for doy 267.
[Adapted from \citet{Dresing-etal-2009}].\label{fig-1}}
\end{figure}
\begin{figure}
\includegraphics[width=\textwidth]{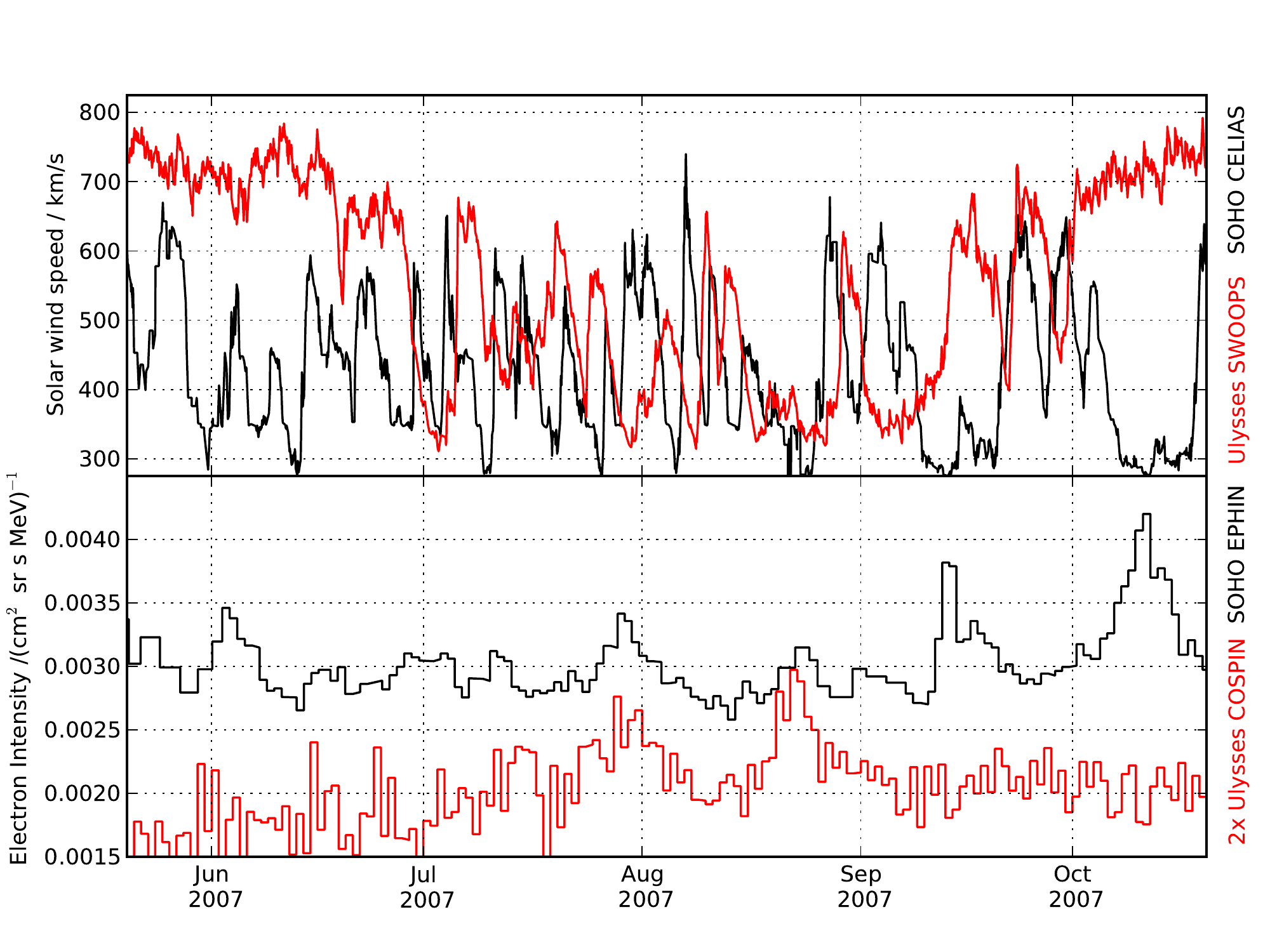}
\caption[]{Ulysses (red curves) and SOHO (black curves) solar wind speed (upper panel) and MeV electron measurements (lower panel) [Adapted from \citet{Kuehl-etal-2013}]. The Ulysses electron intensities have been multiplied by a factor of two.\label{fig-2}}
\end{figure}
A unique constellation of spacecraft to investigate these intensity variation was present in 
August 2007, when Ulysses crossed the heliographic equator during its third so-called fast
latitude scan. Figure~\ref{fig-1} (left) displays the trajectories of Ulysses (blue), STEREO-B
(red), SOHO (gold) and STEREO-A (green) in ecliptic coordinates.  The right panel of that figure 
shows the spacecraft position projected onto the ecliptic plane on September 24. The dotted and
colored lines display Parker spirals using a velocity of 400 km s$^{-1}$.\\
Figure \ref{fig-2} displays the corresponding solar wind and MeV electron measurements by the SWOOPS
\citep{Bame-etal-1992} and SWEPAM \citep{McComas-etal-1998} instruments as well as the MeV electron 
fluxes from the COSPIN/KET and the COSTEP/EPHIN detectors \citep{Mueller-Mellin-etal-1995} aboard
Ulysses, ACE, and SOHO, respectively, for the period of interest. While a recurrent structure of the electron
intensities is present for the whole period shown in Figure~\ref{fig-2}, the Ulysses measurements
only show such variation when the spacecraft is embedded in a CIR region. A first analysis of the
EPHIN measurements has been reported by \citet{Kuehl-etal-2013}, who showed that over half a year of
measurements the electron flux can be correlated or anti-correlated with the solar wind speed
depending on the spacecraft position relative to Jupiter. \citet{Dresing-etal-2009} investigated the spatial and temporal variation of these CIR structures and concluded that the
measurements are highly dependent on the spacecraft latitude and the evolution of the coronal hole
structure. Thus, in order to interpret simultaneous measurements from different locations in the
heliosphere, a detailed knowledge of the plasma background is needed.

\subsection{Background plasma}
The background plasma and magnetic fields in which the transport of energetic particles is modeled can to a first approximation be prescribed by the simple and well-known Parker model \citep{Parker-1958}. Also, attempts have been made to describe CIRs analytically, assuming them to be stationary in a frame corotating with the Sun, e.g.~by \citet{Giacalone-etal-2002}. Such a treatment was recently pursued by \citet{Vogt-2013} to investigate recurrent 27-day decreases in Jovian electron counts \citep[see also][]{Kissmann-etal-2004}.\\
A far more realistic modeling of the heliospheric environment can be obtained with MHD simulations: while, physically, the heliospheric magnetic field (HMF) originates in the Sun, it is conceptually customary to distinguish for modeling purposes between (i) the coronal magnetic field — filling the region from the solar surface out to a spherical so-called heliobase \citep{Zhao-Hoeksema-2010} at several (tens of) solar radii — and (ii) the HMF beyond. There are two popular modeling approaches for the coronal magnetic field, namely potential field reconstructions and MHD models \citep[see][]{Riley-etal-2006}. The latter approach is computationally more challenging but can account for more physics, direct time-dependence and self-consistency. There are numerous examples for such MHD modeling of the coronal magnetic field, including the work of, e.g.,~\citet{Usmanov-Goldstein-2003,Cohen-etal-2007,Lionello-etal-2009,Riley-etal-2011,Feng-etal-2012}.\\
The second popular approach for deriving solar wind conditions utilizes potential field reconstructions of the coronal field that account for the solar wind's influence by introducing a so-called source surface beyond which the field is required to be purely radial. The basic technique of potential field source surface (PFSS) models, originally introduced by \citet{Altschuler-Newkirk-1969} and \citet{Schatten-etal-1969}, is still widely used and was found to provide a means for predicting solar wind speed at Earth via the so-called fluxtube expansion factor of open coronal field lines \citep{Wang-Sheeley-1990}. Another quantity that can be derived from potential field models, the footpoint distance of an open field line to the nearest coronal hole boundary, was used by \citet{Riley-etal-2001} to empirically quantify the resulting solar wind speeds. Combining such approaches and incorporating the Schatten current sheet (SCS) model \citep{Schatten-1971} to account for thin current sheets resulted in the so-called Wang-Sheeley-Arge (WSA) model \citep{Arge-Pizzo-2000,Arge-etal-2003}. A variety of versions of the WSA model predict solar wind speed distributions at different solar distances rather close to the Sun from where the predictions must be propagated further outwards. Earlier models used simple kinematic propagation schemes, while nowadays MHD codes are utilized since they can account for more physics needed, e.g., for the proper modeling of stream interactions and shock formation. Such combined models are in operation at space weather forecasting facilities such as the Community Coordinated Modeling Center (CCMC) or the Space Weather Prediction Center (SWPC).\\
The main advantage of the latter empirical models are their significantly reduced computational costs as compared to coronal MHD models. Additionally, the empirical models avoid the problems arising due to sub-Alfv\'{e}nic solar wind speeds -- complicating boundary conditions as perturbations may travel back towards the photosphere -- and the issue of coronal heating. Furthermore, it was demonstrated by \citet{Riley-etal-2006} that PFSS solutions often closely match those obtained by numerical MHD models.\\
In the present study, we are mainly interested in the influence of CIRs on the transport of energetic particles from distant sources towards the Earth (galactic cosmic rays, Jovian electrons), such that a detailed coronal model is not mandatory. In this light we postpone the implementation of a coronal MHD model to future studies and instead use the empirical WSA model as input to MHD simulations in a domain from 0.1~AU to 10~AU and possibly further out. This will provide a realistic heliospheric environment for the SDE transport code to study propagation of energetic particles, which will be adressed in the second paper of this series. The paper at hand describes the MHD modeling and is organized as follows:\\
Section \ref{sec:code} briefly describes the CRONOS MHD code in the specific application to heliospheric modeling. This setup is validated in Section \ref{sec:pizzo} where we compare results for analytically prescribed boundary conditions for CIRs with those originally obtained by \citet{Pizzo-1982} for the same setup. Section \ref{sec:WSA} gives an overview of the WSA model comprising potential field modeling and a set of empirical formulae to derive the inner boundary conditions for the MHD simulations. The acquisition of data from in situ measurements from the STEREO and Ulysses spacecraft and the comparison with our results is adressed in Section \ref{sec:results} before giving our conclusions and an outlook on future tasks.


\section{Code Setup}
\label{sec:code}
The tool of choice for our simulations is the state-of-the-art MHD code CRONOS, which has been used in recent years to model astrophysical (e.g.~ISM turbulence \citep{Kissmann-etal-2008}, accretion disks \citep{Kissmann-etal-2011}) and heliospherical scenarios \citep{Kleimann-etal-2009, Dalakishvili-etal-2011, Wiengarten-etal-2013}. Amongst its main features the code employs a semi-discrete finite-volume scheme with Runge-Kutta time integration and adaptive time-stepping, allowing for different approximate Riemann solvers. The solenoidality of the magnetic field is ensured via constrained transport, provided the magnetic field is initialized divergence-free. Supported geometries are Cartesian, cylindrical, and spherical (including coordinate singularities) with the option for non-equidistant grids. Here, we use  spherical coordinates ($r$,$\vartheta$,$\varphi$) with the origin being located at the center of the Sun. Thus, $r$ is the heliocentric radial distance, $\vartheta\in[0,\pi]$ is the polar angle (with the north pole corresponding to $\vartheta=0$) and $\varphi\in[0,2\pi]$ is the azimuthal angle. $\varphi$ corresponds to Carrington longitude in this paper, except for the test case in section \ref{sec:pizzo}, where a reference longitude is arbitrary. The code runs in parallel (MPI) and supports the HDF5 output data format.\\
In its basic setup, the code solves the full, time-dependent, normalized equations of ideal MHD
\begin{eqnarray}
  \partial_t \rho + \nabla \cdot (\rho {\bf v}) &=& 0    \\
  \partial_t (\rho {\bf v}) + \nabla \cdot \left[\rho {\bf v v} 
    +  (p + |{\bf B}|^2/2) \ {\bf 1} - {\bf B B} \right]
  &=& {\bf f} \label{eq:mom}\\
  \partial_t e + \nabla \cdot \left[ (e+p+ |{\bf B}|^2/2) \ {\bf v}
    - ({\bf v} \cdot {\bf B}) {\bf B} \right] &=& {\bf v}\cdot{\bf f}  \\
   \partial_t {\bf B} + \nabla \times {\bf E}  &=& {\bf 0} 
\end{eqnarray}
where $\rho$ is the mass density, ${\bf v}$ is the fluid velocity, ${\bf B}$ and ${\bf E}$ describe the electromagnetic field, $e$ is the total energy density, and $p$ is the scalar thermal pressure.\\
Additional force densities ${\bf f}$ can be introduced by the user. In our setup these are the gravitational force density ${\bf f}_g = -\rho GM_\odot/r^2{\bf e}_r$ and, since it is convenient to perform calculations in a frame of reference corotating with the Sun, the fictitious forces ${\bf f}_{cor} = - 2\rho{\bf\Omega}\times{\bf v} -  \rho{\bf\Omega}\times({\bf\Omega}\times{\bf r})$, where ${\bf \Omega} = \Omega {\bf e}_z$. Since we know of no consistent way to connect the Sun's observed differential surface rotation from the photosphere to the lower radial boundary of our computational domain, we are forced to neglect this effect and, therefore, choose a constant solar angular rotation speed $\Omega = 14.71^\circ/{\rm d}$ \citep{Snodgrass-Ulrich-1990}. Furthermore, ${\bf 1}$ denotes the unit tensor, and the dyadic product is used in (\ref{eq:mom}).\\
Amongst the closure relations
\begin{eqnarray}
  e &=&
  \frac{\rho \ |{\bf v}|^2}{2} +
  \frac{       |{\bf B}|^2}{2} +
  \frac{p}{\gamma-1} \\
  {\bf E} + {\bf v} \times {\bf B} &=& {\bf 0} \\
  \nabla \cdot {\bf B} &=& 0 ~, 
\end{eqnarray}
an adiabatic equation of state is used with an adiabatic exponent $\gamma=1.5$, which is the same value as used in the heliospheric part of the ENLIL setup \citep{Odstrcil-etal-2004}.\\

\section{Model Validation}
\label{sec:pizzo}
As mentioned in the introduction, the modeled heliospheric background will be used as an input for a SDE transport code in order to investigate the propagation of energetic particles in a forthcoming study. During solar quiet times, CIRs are the most prominent agents that can have significant influence on the transport coefficients as they act as diffusion barriers due to the associated magnetic field enhancements. We, therefore, first demonstrate the capability of our setup to model these structures. Further detailed model validation for the application of the CRONOS code to inner-heliospheric scenarios were performed by \citet{Wiengarten-etal-2013}.\\
There are no exact analytic expressions for the plasma quantities in CIR associated structures, although there exist simplified models (e.g. \citet{Giacalone-etal-2002}, see also \citet{Vogt-2013}), but the expressions given therein perform a heuristical fit to data and lack a description of the CIR associated shocks. Therefore, to validate our model we instead compare with results obtained with previous numerical simulations, namely the pioneering work by \citet{Pizzo-1982} who investigated a variety of different steady-state scenarios, one of which will be summarized and compared with here.
\begin{figure}
\begin{center}
\includegraphics[width=\textwidth]{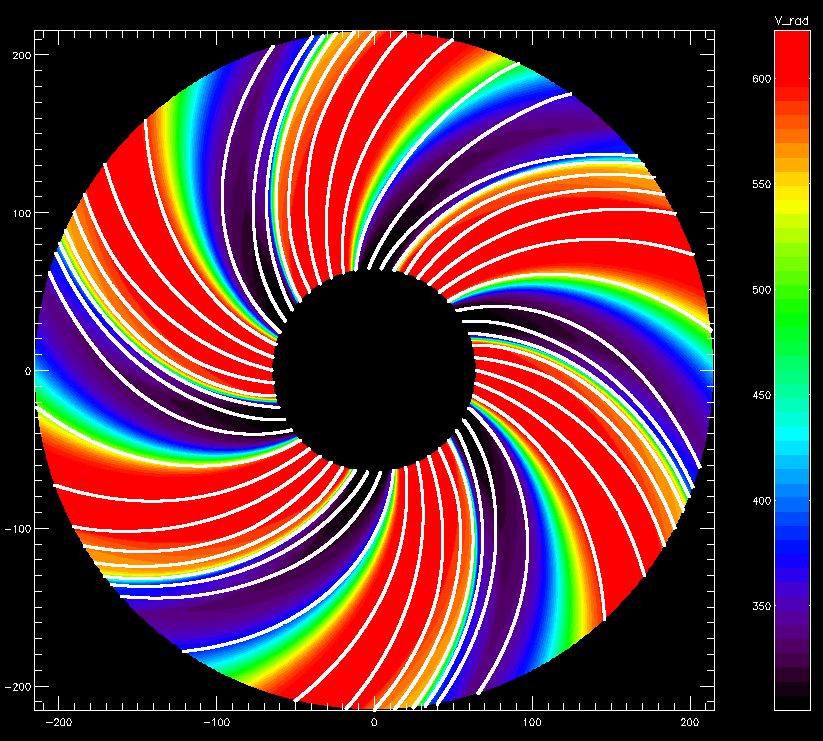}
\caption{Visualization of the modeling test case in an equatorial slice: Color coding refers to radial velocity, magnetic fieldlines are shown in white.}
\label{fig:pizzo_overview}
\end{center}
\end{figure}
The inner radial grid boundary is chosen to be located at $r_0=0.3$~AU, well beyond the critical surfaces, so that the solar wind speed is super-magnetoacoustic everywhere. Therefore, perturbations cannot travel towards the Sun and constant (corotating) boundary conditions can be chosen. A circular regional patch of fast, tenuous and hot wind centered at the equator is embedded into the ambient slow, dense and cool wind, providing the basic ingredients for a CIR (see Figure \ref{fig:pizzo_overview} to get a first expression). The inner radial boundary conditions are shown in an equatorial slice in Figure \ref{fig:pizzo_slice} (a) and are ascribed as follows. The formula used to specify radial velocity at the inner boundary in our setup reads $v(r_0,\vartheta,\varphi) = v_0(1+a)$ with $v_0=300$~km/s and $a\in[0,1]$, so that the ambient slow velocity is 300~km/s and the fast wind is at 600~km/s. Specifically, 
\begin{equation}
a=\sum_{m=1}^{M} 0.5[\tanh(A(\xi_m+\Delta\xi))-\tanh(A(\xi_m-\Delta\xi))-\tanh(A\xi_m)]
\label{eq:pizzo}
\end{equation}
where $\xi_m=\sqrt{(\vartheta-\vartheta_0)^2+(\varphi-\varphi_{0,m})^2}$ is the angular distance from the center of the high-speed stream at $(\vartheta_0,\varphi_{0,m})$, $\Delta\xi$ is its angular extent, and $A$ controls the steepness of the transition from slow to fast wind. This approach was necessary since there is no explicit formula given by \citet{Pizzo-1982}, but instead the initial configuration shown (their Figure 5) was digitalized (http://digitizer.sourceforge.net/) and fitted with parameters for the above formula resulting in $A=30$, $\Delta\xi=13^\circ$, $\vartheta_0=\pi/2$, $\varphi_{0,m}=(2m-1)\pi/M$. The upper index $M$ in the sum of Equation (\ref{eq:pizzo}) is set to $M=6$ in accordance with the Pizzo setup, which refers to the fact that there are actually six high-speed streams in the whole $2\pi$ interval, giving a periodicity of $60^\circ$, so that $\varphi_{0,m}$ gives the longitudinal center of the respective high-speed stream. Other values for $M$ could be chosen to investigate the interaction of different CIRs depending on their longitudinal separation.\\
Density is inversely correlated to velocity via $n(r_0,\vartheta,\varphi)=n_0(1+a)^{-1.5}$ with $n_0=120~{\rm cm}^{-3}$ so that with a prescribed constant pressure, temperature is inversely proportional to $n$ with a value $T_0=0.16$~MK for the ambient wind. For the magnetic field strength, \citet{Pizzo-1982} assumes a constant value of $B_0=45$~nT. In our setup we prescribe a constant $B_{r,0}$ and calculate $B_{\varphi,0}=-B_{r,0}v_{\varphi,0}/v_{r,0}$ so there is a small deviation from a constant $B_0$ since $v_{r,0}$ varies as described above. Furthermore, we assume a small value for the azimuthal velocity component $u_{\varphi,0}=\Omega (R_c/r_0) \sin(\vartheta)$ in the inertial frame with a radius of corotation $R_c=1.5R_\odot$, as follows from the Weber-Davis model \citep{Weber-Davis-1967} also considered in our previous work \citep{Wiengarten-etal-2013}. The azimuthal velocity in the corotating frame then is $v_{\varphi,0}=u_{\varphi,0}-\Omega r_0\sin(\vartheta)$.
\begin{figure}
\includegraphics[width=\textwidth]{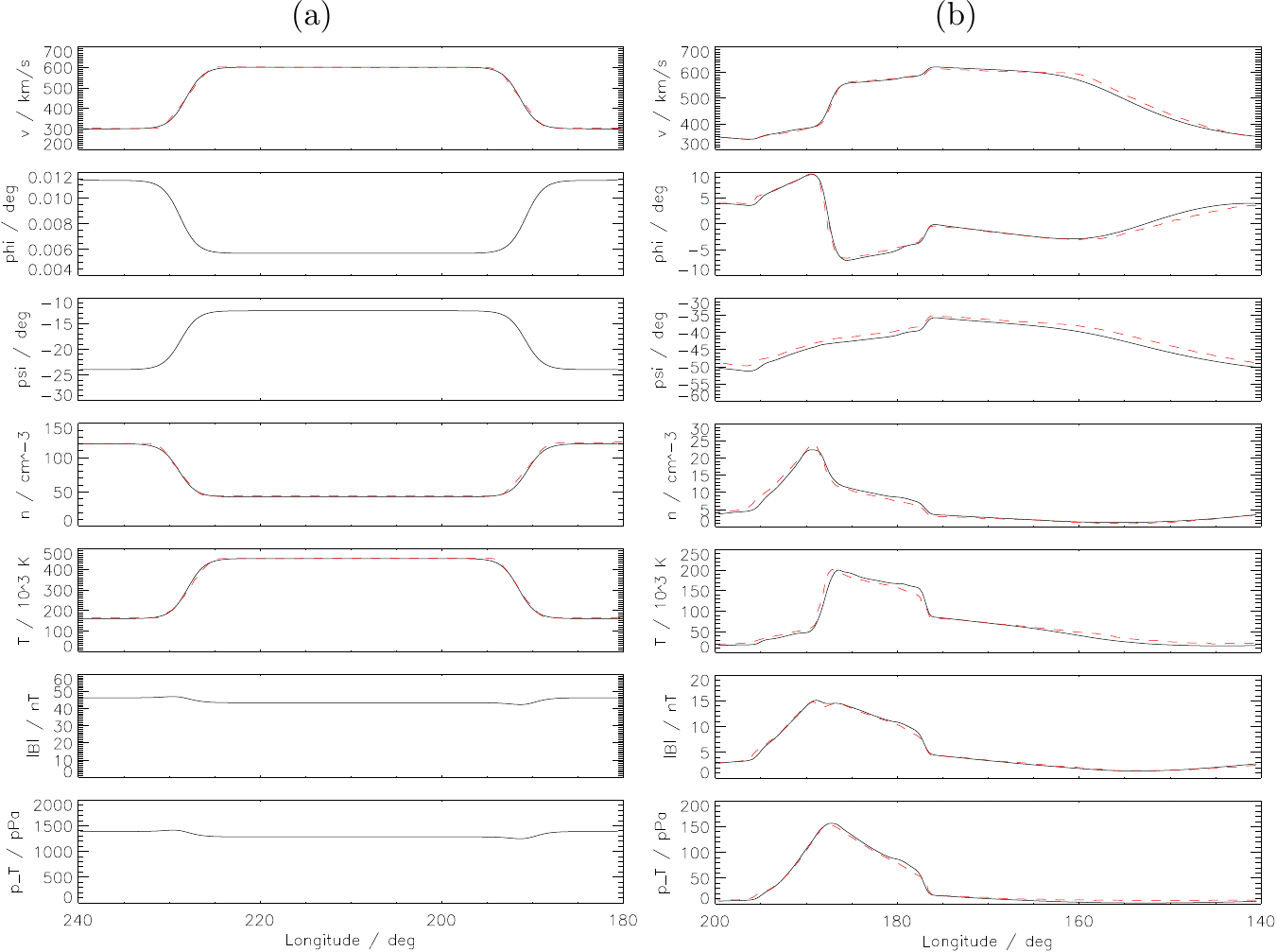}
\caption{Comparison with Pizzo for (left) initial conditions (at 0.3~AU) and (right) 1 AU results in the equatorial plane. Shown quantities are (top to bottom): radial velocity $v_r$, flow angle in corotating frame $\phi$ and in inertial frame $\psi$, proton number density $n$, temperature $T$, magnetic field strength $B$, and total pressure $p$. CRONOS data are in black-solid, Pizzo data in red-dashed.}
\label{fig:pizzo_slice}
\end{figure}
The longitudinal periodicity, giving a total of six high-speed streams in the whole 2$\pi$ interval, allows for a rather high angular resolution ($\Delta\varphi=\Delta\vartheta=0.25^\circ$): the computational costs are kept low by performing the calculations for just one embedded structure and applying periodic boundary conditions in the azimuthal direction, so that the azimuthal extent of the simulation box $\varphi\in[0,2\pi/6]$. For the results presented below, the simulation results have been copied and appended to cover the whole $2\pi$ interval. The polar regions are not significant here and are excluded to avoid small time-steps, thus $\vartheta\in[0.1\pi,0.9\pi]$. In the radial direction the simulation box extends to 10~AU, with the radial cell size set to $1R_\odot$ on a linear grid. Computations are performed until a steady state is reached. A 2D visualization is shown in an equatorial slice (for the whole longitudinal interval, restricted to 1AU) in Figure \ref{fig:pizzo_overview} with the color coding according to radial velocity. The density distribution of the magnetic field lines indicates the formation of compression and rarefaction regions. Quantitative results at 1~AU are shown in Figure \ref{fig:pizzo_slice} (b) in the same manner as the inner boundary conditions, where the results to compare with have been digitized as above from Figure 6 in \citet{Pizzo-1982}. The agreement between our results and Pizzo's is very satisfactory, bearing in mind that on the one hand there are slight differences in boundary conditions and on the other hand different codes and numerical schemes were used. In fact, Pizzo solved the steady-state MHD equations, while the full time-dependent set of equations is treated here.
\begin{figure}
\begin{center}
\includegraphics[width=\textwidth]{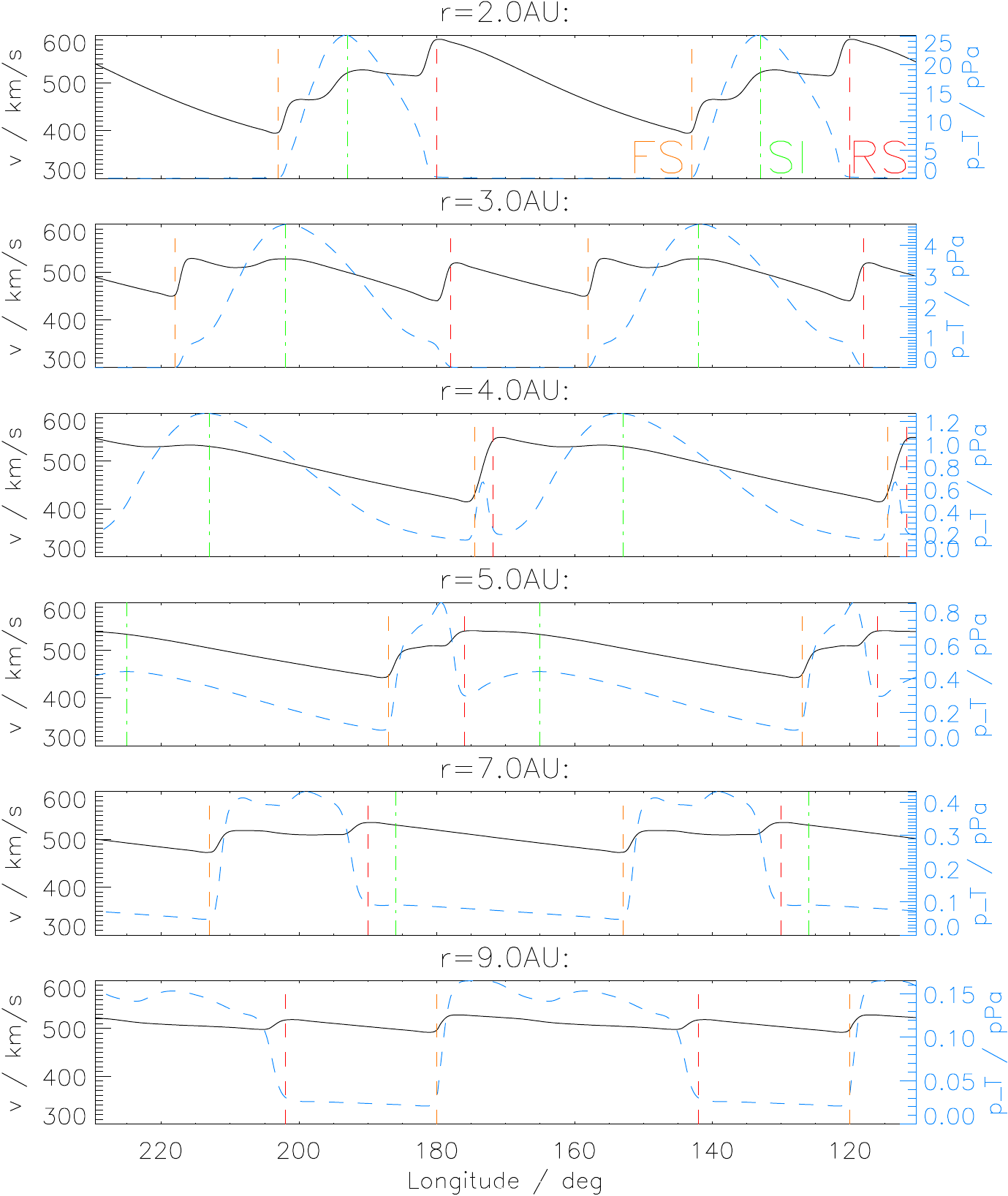}
\caption{Evolution and interaction of two adjacent CIR structures in the equatorial plane from 2~AU to 9~AU. Radial velocity is shown in solid black with a fixed y-axis scaling, total pressure in dashed blue with adaptive y-axis scaling. The longitudinal positions of the stream interface (SI, dashed-dotted green), forward shock (FS, dashed orange) and reverse shock (RS, dashed red) are depicted by respective vertical lines. See text for details.} 
\label{fig:pizzo_evolution}
\end{center}
\end{figure}
Compression occurs at the leading edge of the high-speed stream at a longitude of $\varphi\approx187^\circ$ with respective elevations in the pressure associated quantities in the bottom panels. There are two waves propagating away from this interface in opposite directions, one forward ($\varphi\approx196^\circ$) and one backward ($\varphi\approx177^\circ$), which have already steepened into MHD shocks. This is the result of the initial sharp transition between the streams, causing a high compression. For different initial conditions shocks may evolve only beyond several AU of radial distance or not at all. The flow angle (second panel) shows that the interface is a region of shear, with azimuthal speeds directing away from the interface, which causes the latter to broaden and smear out with increasing distance from the Sun. Further steepening of the forward-reverse shock pair and a broadening of the interaction region occur for larger radial distances.\\
This is explicitly visible in Figure \ref{fig:pizzo_evolution}, depicting the evolution and interaction of two adjacent structures out to a radial distance of 9~AU. Here, radial velocity is shown in solid black with a fixed y-axis scaling, while total pressure shows as dashed blue with adaptive y-axis scaling to account for the rapidly decreasing values. The longitudinal interval shown in all panels is extended to $120^\circ$ and is fixed in the arbitrary range from 110$^\circ$ to $230^\circ$. At 2~AU the forward (FS) and reverse (RS) shocks are more pronounced and are indicated via horizontal dashed lines in orange and red, respectively. The stream interface (SI) is identified as local maxima in total pressure, which coincides with the zero-crossing of the flow angle $\phi$ (not shown here). The SI is indicated as green dashed-dotted vertical line. Note, for the following panels, that it is not the same two CIRs shown in the consecutive panels: due to the Parker angle of $45^\circ$/AU, the structures experience a respective shift in longitude to the right from one panel to the next one, while the periodicity of $60^\circ$ gives the impression of structures apparently moving to the left. For the SI this gives exactly the difference of $15^\circ$ to the left per panel, while the shocks propagate away from it. They encounter each other just before 4~AU and a corotating merged interaction region (CMIR) is formed. The shocks move through each other and propagate further at a lower relative speed, due to energy losses in the collision process \citep[p.110]{Parker-1963}. This gives rise to another compression region that is more pronounced than the original one at the SI by 5~AU and it continues to be the dominant structure out to 9~AU, while the shocks weaken to pressure waves. This demonstrates the different dynamics: In the inner heliosphere it is momentum driven by streams, while beyond several AU it is driven by the evolution and interaction of interaction regions and shocks \citep[p.137]{Burlaga-1995}.   

This rather simple setup will also provide a good first test for the SDE transport modeling, because the relevant structures are regular and fairly easy to identify.  

\section{Utilizing Observational Input Data}
\label{sec:WSA}
Having validated the numerical framework, we now seek to model more realistic solar wind conditions as present during a given period of time. With a focus on CIRs, we chose a time period in late 2007 (Carrington Rotations (CR) 2059 to 2061) when, due to coronal hole excursions to low latitudes, several fast wind streams were present in the otherwise slow ambient solar wind. Because there were no transient events, this time period is also favorable for the present study as the used WSA model does not cover CMEs. The WSA model relies on the topology of PFSS models of the coronal magnetic field. To obtain the PFSS solution, we used magnetograms obtained by the Global Oscillation Network Group (GONG) of the National Solar Observatory (NSO) (http://gong.nso.edu/) as input to the "Finite Difference Iterative Poisson Solver" (FDIPS, \citet{Toth-etal-2011}). The two following sections summarize our implementation of the WSA model using the example of CR 2060.

\subsection{Potential Field Modeling} 
PFSS models \citep{Altschuler-Newkirk-1969, Schatten-etal-1969} assume a current-free corona, so that from $\nabla\times {\bf B}={\bf 0}$ the magnetic field can be expressed in terms of a scalar potential (${\bf B}=-\nabla\psi$) that, due to the solenoidality condition $\nabla\cdot{\bf B}=0$, has to fulfill Laplace's equation $\nabla\cdot(\nabla\psi)=0$. To simulate the effect that the solar wind has on the magnetic field, which is to drag it out with increasing radial distance, a source surface is introduced (typically at $R_{s}=2.5R_\odot$) beyond which the field is assumed to be purely radial. Laplace's equation is commonly solved by series of spherical harmonics. While this provides a computationally cheap way to describe the large-scale coronal magnetic field, an inherent disadvantage of the spherical harmonics approach is the occurence of ringing artifacts around sharp features when the order of spherical harmonics approaches the resolution of the magnetogram \citep{Tran-2009}. An alternative approach using a finite difference scheme to solve Laplace's equation was proposed by \citet{Toth-etal-2011}, and their code FDIPS was made publicly available (http://csem.engin.umich.edu/tools/FDIPS/). Their approach completely avoids artifacts and the magnetogram is exactly reproduced. The computational costs to obtain the potential field solution are also fairly low on modern computers (i.e.~minutes when using the parallelized version on a 16 core machine). We use FDIPS with GONG integral synoptic magnetograms in the Fits-Format and apply a resolution of $[N_r,N_\vartheta,N_\varphi]=[150,180,360]$ grid points, so that the angular resolution matches that of the magnetograms. The resulting grid's radial resolution $\Delta r=0.01R_\odot$ agrees with the stepsize in tracing magnetic field lines in the WSA model (C.~N.~Arge, personal communication).
\begin{figure}
\includegraphics[width=\textwidth]{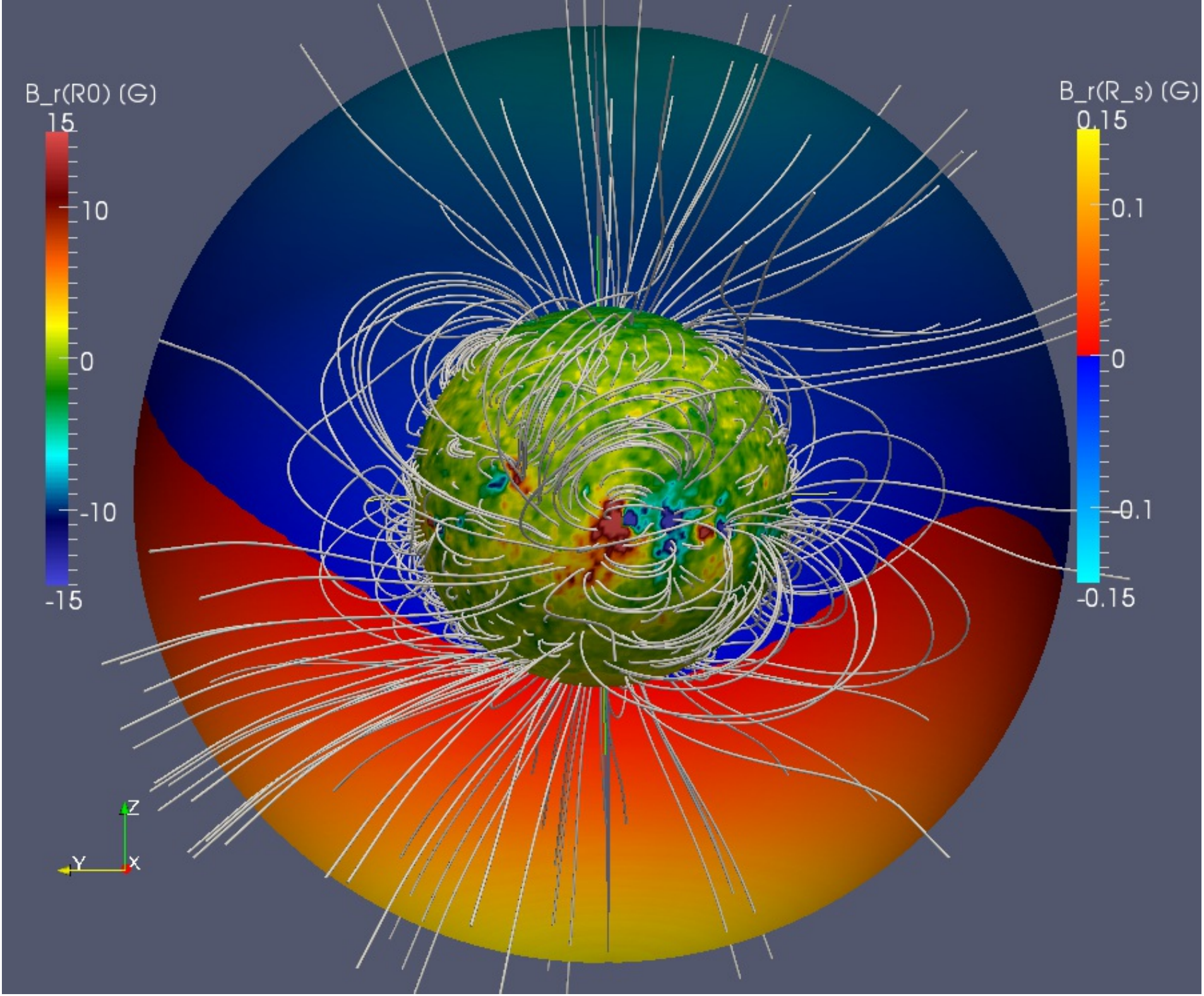}
\caption{PFSS configuration for CR2060: the photospheric magnetogram is shown on the inner sphere, where the data range is restricted to $\pm15$ Gauss. The outer half-sphere represents the source surface. Open magnetic field lines reach the source surface and are mainly confined to polar regions; however, excursions to lower latitudes are visible as well.}
\label{fig:pfss}
\end{figure}
The resulting potential field configuration for CR2060 is shown in Figure \ref{fig:pfss}: the reconstructed photospheric magnetogram is shown on the inner sphere (see top panel of Figure \ref{fig:cs} for a full map), where the data range is restricted to $\pm15$ Gauss to better illustrate the small scale structures (active regions in this magnetogram have maximum values up to 500 Gauss). The outer half-sphere represents the source surface. Selected field lines display rather typical solar minimum conditions with open field lines in polar regions (giving rise to the fast polar wind) and closed loops towards the ecliptic plane (resulting in a slow wind there). Excursions of open field lines to lower latitudes give rise to high-speed streams there that will interact with the ambient slow wind to form CIRs.\\
\begin{figure}
\centering
\includegraphics[scale=0.35,angle=0]{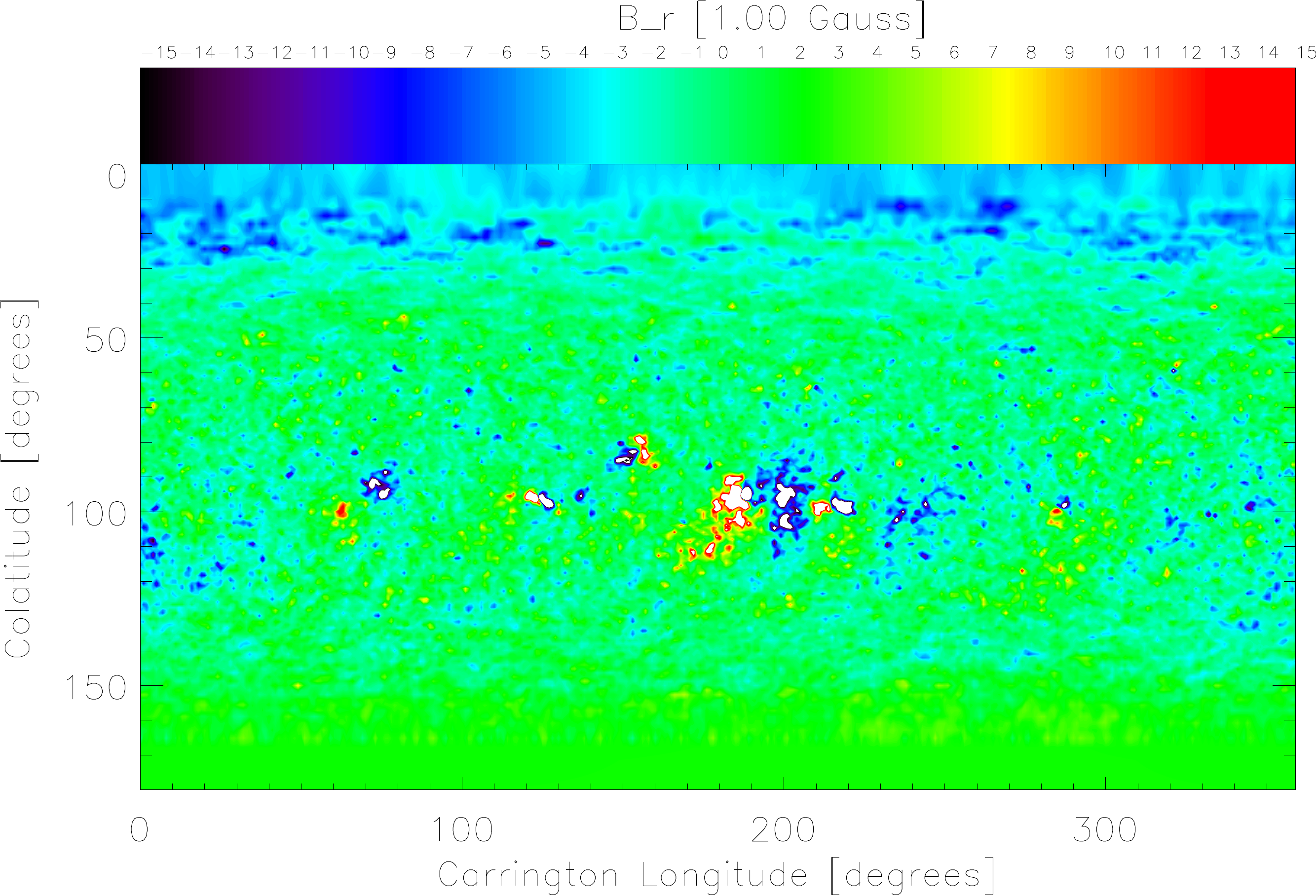}
\hfill
\includegraphics[scale=0.35,angle=0]{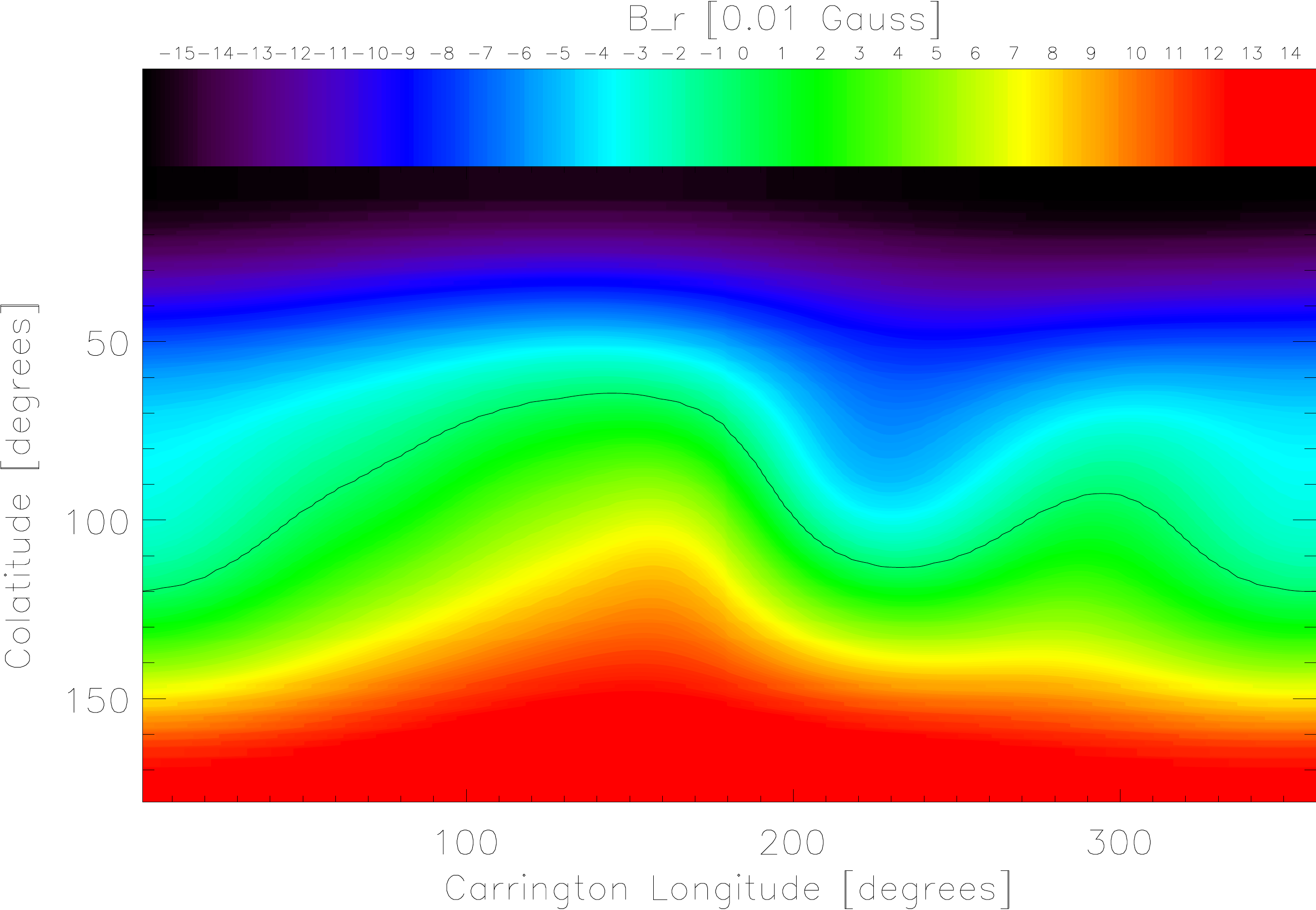}
\hfill
\includegraphics[scale=0.35,angle=0]{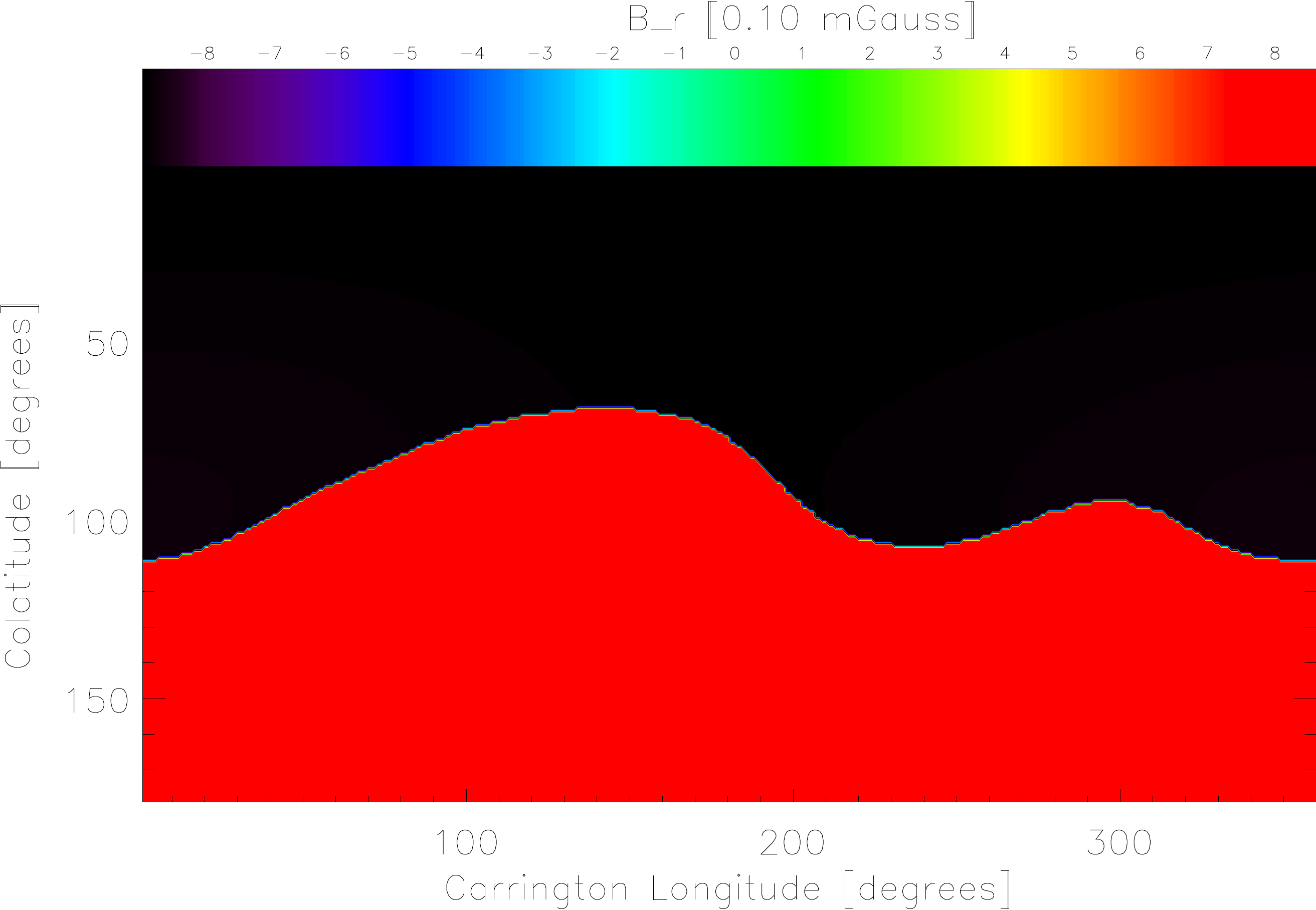}
\hfill
\caption{Maps of radial magnetic field strength, top: at the photosphere, middle: at the source surface, bottom: at the inner radial MHD grid boundary from SCS model.}
\label{fig:cs}
\end{figure}
It has been demonstrated that PFSS solutions often closely match respective MHD results \citep{Riley-etal-2006}. One shortcoming of the PFSS models, however, is that they do not produce a thin current-sheet, which can be seen in Figure \ref{fig:cs}, middle panel, showing a map of the source surface radial magnetic field where the transition between the different polarities is rather broad. This is in contrast to observations that show sharp and thin current sheets at magnetic field polarity reversals. To overcome this problem, the WSA model further utilizes the Schatten current sheet (SCS) model \citep{Schatten-1971} to compute the magnetic field beyond the source surface: the magnetic field at the source surface is first re-orientated where necessary to point (radially) outward everywhere, and is then used as a boundary condition for another potential field approach from which respective spherical harmonic coefficients are calculated. In contrast to the reconstruction of the highly structured magnetograms, the spherical harmonics approach does not suffer from ringing artifacts in this case, since a small maximal order of 9 is sufficient and applied here. The initial orientation has to be restored in the resulting configuration, which can be achieved by tracing field lines from the outer boundary of the Schatten model at $R_{gb}=0.1$~AU back to the source surface. The resulting map of the radial magnetic field (Figure \ref{fig:cs}, bottom panel) at $R_{gb}$ is topologically similar to the one at the source surface with smaller maximal tilt angles of the current sheet. The map is used in defining the inner boundary condition in the MHD calculations as described in the next section.

\subsection{Empirical Interface} 
\begin{figure}
\plotone{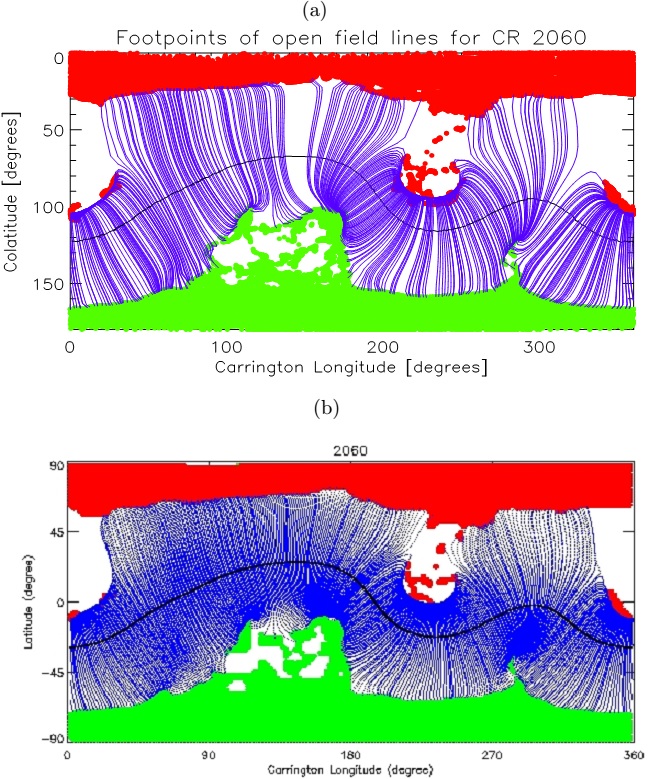}
\caption{Footpoint locations of open field lines (red/green dots) and highest closed fieldlines (blue) in (a) our setup and (b) as taken from the GONG website for comparison.}
\label{fig:footpoints}
\end{figure}

To derive boundary conditions for the remaining plasma quantities at $R_{gb}=0.1$~AU, a set of empirial formulas is employed, which are largely based on the topology of the coronal potential field configuration. \citet{Wang-Sheeley-1990} found an inverse relationship between the flux-tube expansion factor 
\begin{equation}
f_s(\vartheta,\varphi) = \left(\frac{R_\odot}{R_{s}}\right)^2\cdot\frac{B(R_\odot,\vartheta_0,\varphi_0)}{B(R_{s},\vartheta,\varphi)}
\label{eq:fs}
\end{equation}
and the resulting solar wind speed: large expansion factors (low solar wind speeds) are usually associated with field lines that have their photospheric footpoints at coronal hole boundaries (i.e. the boundary between open and closed field lines), while small expansion factors (high solar wind speeds) are associated with those originating from deep within a coronal hole. Therefore, the further away an open field line footpoint is located from a coronal hole boundary the higher is the resulting solar wind speed. This can be expressed in terms of the quantity $\theta_b$, the footpoint distance to the nearest coronal hole boundary, introduced first by \citet{Riley-etal-2001}. It was since then found that using both $f_s$ and $\theta_b$ in conjunction gives better results than either one alone \citep{Arge-etal-2003}.\\
The actual computation of $f_s$ and $\theta_b$ requires field lines being traced back to their respective footpoints in the photosphere at ($\vartheta_0,\varphi_0$). Our algorithm implemented for tracing the field lines employs an adaptive step-size method inherent to embedded Runge-Kutta (RK) methods \citep{Press-etal-2007}, where in our setup the maximal allowable deviation $D_{RK}$ from unity ratio taken of 5th order RK and embedded 4th order RK is used to determine the step-size. For the tracing in the PFSS domain below $2.5R_\odot$ we used $D_{RK}=10^{-3}$ and $D_{RK}=10^{-4}$ below $1.1R_\odot$, respectively, resulting in stepsizes in the range from 0.1 to 0.01$R_\odot$ (with the lower limit corresponding to the grid's cellsize). A comparison using $D_{RK}=10^{-4}$ everywhere yielded no difference in resulting footpoint locations within $0.01^\circ$, which is far below the magnetograms resolution. Similarly, for the SCS models domain beyond 2.5$R_\odot$ step-sizes can go up as high as 0.8$R_\odot$ for $D_{RK}=10^{-3}$. The resulting photospheric footpoint locations are shown in Figure \ref{fig:footpoints} (a): The red/green dots indicate footpoints with positive/negative polarity, respectively. Besides the large coronal holes in the polar regions there are excursions to equatorial latitudes, which are the sources of respective high-speed streams there. Also shown are the highest closed fieldlines in blue, which are traced in both directions from just below the source surface ($r_b=2.49R_\odot$) and characterized as closed if the photosphere is reached in both directions. A qualitative comparison can be made with similar plots available on the GONG website\footnote{\url{http://gong.nso.edu/data/magmap/QR/mqf/200708/mrmqf070828/mrmqf070828t0501c2060\_000.gif}}, which is shown for CR2060 in panel (b) of Figure \ref{fig:footpoints}. Even though there are slight differences in the details, the global topology is very similar to that in panel (a), especially concerning the locations of the equatorial extensions of open fieldline footpoints and regions with fieldlines closing below $r_b$. Differences may arise due to the different potential field approaches (the plot from the GONG webpage uses spherical harmonics) and the method to calculate the field lines.\\
The coronal hole boundary can now be defined to be situated where footpoints of open field lines are adjacent to those of closed field lines. This was achieved by binning the respective photospheric map into a $1^\circ \times 1^\circ$ grid and then labeling cells based on whether or not they contain open field lines. The coronal hole boundary is defined where an 'open cell' has at least three of its eight neighboring cells marked as 'closed'. We calculate the distance to the nearest coronal hole boundary (CHB) for each footpoint of an open field line (FP) by taking the distance 
\begin{eqnarray}
d = \arccos&(&\sin(\vartheta_{CHB})\cos(\varphi_{CHB})\sin(\vartheta_{FP})\cos(\varphi_{FP}) \nonumber \\
					&+& \sin(\vartheta_{CHB})\sin(\varphi_{CHB})\sin(\vartheta_{FP})\sin(\varphi_{FP}) \nonumber \\
					&+& \cos(\vartheta_{CHB})\cos(\vartheta_{FP})) 
\end{eqnarray}
along a great circle to all cells marked as coronal hole boundary and choosing the smallest value.\\
The computation of the fluxtube expansion factor $f_s$ is carried out along with the determination of open footpoints by taking the respective magnetic field values at the source surface and the photosphere as connected by a field line and using Formula (\ref{eq:fs}).\\ 
The set of empirical formulae to determine the boundary and initial conditions for the remaining MHD quantities is similar to the one used by \citet{Wiengarten-etal-2013}, with the following adaptions:\\ the formula for radial velocity now reads
\begin{equation}
\label{eq:mcgregor}
v_r = V_0 + \frac{V_1}{(1+f_s)^{2/9}}\left(1.0-0.8\exp\left(-\left(\frac{\theta_b}{\phi}\right)^{\beta}\right)\right)^3 \ ~,
\end{equation}
as discussed by \citet{McGregor-etal-2011}. A number of such formulae can be found in the literature \citep[e.g.][]{Feng-etal-2010, Detman-etal-2011}, differing in form as well as in scaling parameters. The parameters have values $V_0=240$~km/s, $V_1=675$~km/s, $\beta=1.25$ and $\phi=2.8^\circ$ at SWPC while \citet{McGregor-etal-2011} found $V_0=200$~km/s, $V_1=750$~km/s, $\beta=3.6$ and $\phi=3.8^\circ$ by fitting solar wind velocity distributions. We performed a rough tuning to be in better agreement with the spacecraft data and found for our setup more suitable values $V_0=200$~km/s, $V_1=675$~km/s, $\beta=2.0$ and $\phi\in[2.8^\circ,3.2^\circ]$, which are comparable to the values listed above. A more thorough tuning would go beyond the scope of this paper, and optimal parameters may vary for different CRs.\\
Since potential field models systematically underestimate the magnetic flux \citep{Stevens-etal-2012}, we apply a correction to the initial magnetic field for better accordance with 1~AU data while keeping the orientation of $B_{gb}$. Specifically, a value of 300~nT is applied for a respective solar wind speed of 625~km/s and linearly scaled for other speeds, as is done in the WSA-Enlil model \citep{McGregor-etal-2011}. Additional estimates for mass flux and total pressure are obtained from OMNIweb (\url{http://omniweb.gsfc.nasa.gov/}) data by taking respective 27-day averaged values. These are then scaled with radial distance to the inner grid boundary and used for defining boundary conditions for density and temperature as described in \citet{Wiengarten-etal-2013}.\\
The derived boundary conditions are assumed to remain stationary in the corotating frame during the evolution of a single CR so that the simulations can be advanced in time until a steady state is reached. The simulation box extends to 2~AU in order to include Ulysses at $r\approx(1.4-1.5)$~AU during the time period considered, and also to study the formation of CIR-associated shocks. The applied resolution is $\Delta r=2R_\odot,\Delta\vartheta=1^\circ,\Delta\varphi=1^\circ$, and, since we focus on the validation of the results with spacecraft data, the simulation box is restricted to extend to 2~AU only in the radial direction and to $\vartheta\in[0.2\pi,0.8\pi]$ in latitude. This gives a runtime $T_{run}\approx15$ hours on a 64 core cluster, while the physical convergence time $T_{conv}\approx300$ hours, which is estimated from the slowest wind ($v_s\approx250$km/s) to propagate to the outer boundary. Taking the simulation to larger radial distances for usage in the SDE code is straightforward, but requires the grid to be coarsened to maintain reasonable required computer resources. The polar regions could be included as well, but this would further restrict the global timestep in the simulations due to small cell sizes as $\sin(\vartheta)\rightarrow0$ towards the poles.\\

\section{Results -- Comparison with Spacecraft Data}
\label{sec:results}
\begin{figure}
\plotone{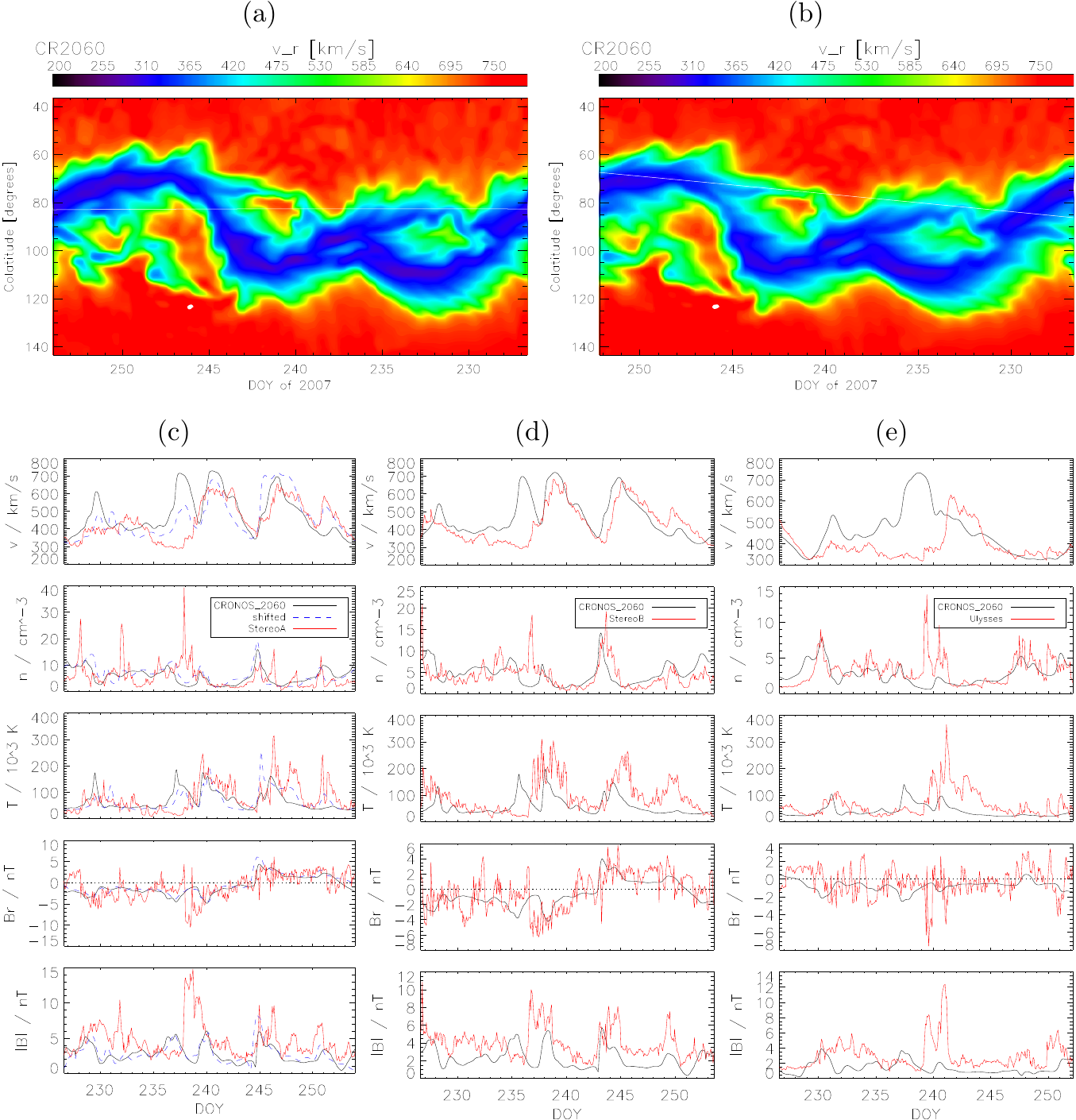}
\caption{Top: Maps for radial velocity at respective spacecraft's radial distance (a) with indicated trajectory of STEREO-A, and (b) for Ulysses. Bottom: Quantitative comparison with spacecraft data: (c) STEREO-A, (d) STEREO-B, (e) Ulysses. Black lines are simulated, red lines spacecraft data and the dashed blue line in (c) refers to a slightly shifted trajectory (see text). The dotted horizontal for $B_r=0$ is to guide the eye in identifying sector boundaries.}
\label{fig:comp}
\end{figure}

The simulation results of the solar wind speed, density, and temperature as well as of the magnetic field are compared to spacecraft data of both the Ulysses and the two STEREO spacecraft. Stereo-A(head) leads Earth while Stereo-B(ehind) trails Earth.
In detail, the Ulysses data are based on measurements from the Ulysses Solar Wind Observations Over the Poles of the Sun (SWOOPS, \citet{Bame-etal-1992}) and the magnetometer onboard (VHM, \citet{Balogh-etal-1992}). For the STEREO-A/B twin spacecraft, measurements were taken from the Plasma and Suprathermal Ion Composition (PLASTIC, \citet{Galvin-etal-2008}) instrument and the spacecraft's magnetometers (MAG, \citet{Acuna-etal-2008}).\\
We first discuss the results for CR 2060 and only briefly address the results for CR2059 and CR2061 presented afterwards.\\
For a comparison with the spacecraft data the MHD results have been interpolated along respective spacecraft trajectories. Panel (a) in Figure \ref{fig:comp} shows the resulting map for the radial velocity at the radial distance of the STEREO-A spacecraft (at $\approx1$ AU and separated in longitude from Earth by less than 15$^\circ$, see \url{http://stereo-ssc.nascom.nasa.gov/where.shtml}). The white line indicates its trajectory as the spacecraft traverses from right to left as indicated by the corresponding day of year (DOY) used as the horizontal axis. A respective map for STEREO-B is very similar to panel (a) and is, therefore, omitted. Instead panel (b) shows the equivalent map for the Ulysses spacecraft, which at that time performed a fast-latitude scan and was located at heliocentric distances of $r\approx(1.4-1.5)$~AU. Quantitative comparisons with the in situ measurements are presented in the bottom panels (c)-(e), where the quantities shown are (top to bottom) radial velocity $v$, particle number density $n$, temperature $T$, radial magnetic field $B_r$, and magnetic field magnitude $B$. The spacecraft data (red lines) have been averaged to the one degree angular resolution of the simulated data (black lines).
A quite good match is found along the orbit of STEREO-B (panel (d)), where the prominent high-speed streams centered around DOYs 239 and 246 are captured in terms of magnitude and stream width, while a third high-speed stream at DOY 250 is somewhat underestimated. Furthermore, the simulation data shows an additional feature at DOY 236, which is not present in the spacecraft data, and can be identified as an excursion of the northern coronal hole in panel (a). A similar behavior is found for STEREO-A. Here we demonstrate the effect of artificially shifting the spacecraft position slightly (4$^\circ$ south in latitude) which results in the dashed blue curve. The comparison seems significantly improved as the additional features prominence is mitigated while the observed stream at DOY 250 is now captured very acurately. This shows that a comparison strictly along a spacecraft trajectory may not always be satisfactory at first glance, but that a thorough inspection of such maps as presented in panels (a) and (b) can help identifying the actually observed ones, which may just be slightly off the trajectory in the simulation (see Section \ref{sec:conclusions} for further discussion). The comparison with Ulysses data for radial velocity is also somewhat dominated by the additional feature around DOY 239, however, apart from that the comparison is quite satisfactory throughout the course of this CR.\\
The pressure associated quantities $n$, $T$, and $B$ exhibit magnitudes of the correct order with respective strong enhancements in compression regions associated with the forming CIRs at the leading edges of high-speed streams. The magnetic field strength, however, seems to be systematically underestimated and might have to be increased in future simulations.\\
Another interesting quantity is the radial magnetic field component, through which sector boundaries (i.e.~current sheet crossings) can be identified, which is difficult, however, due to rapid fluctuations in the spacecraft data. Still, the average polarity and sector boundaries are captured fairly well, e.g.~for STEREO-B, current sheet crossings occur at DOY 243 and 250 in good agreement with the measurements.
\begin{figure}
\plotone{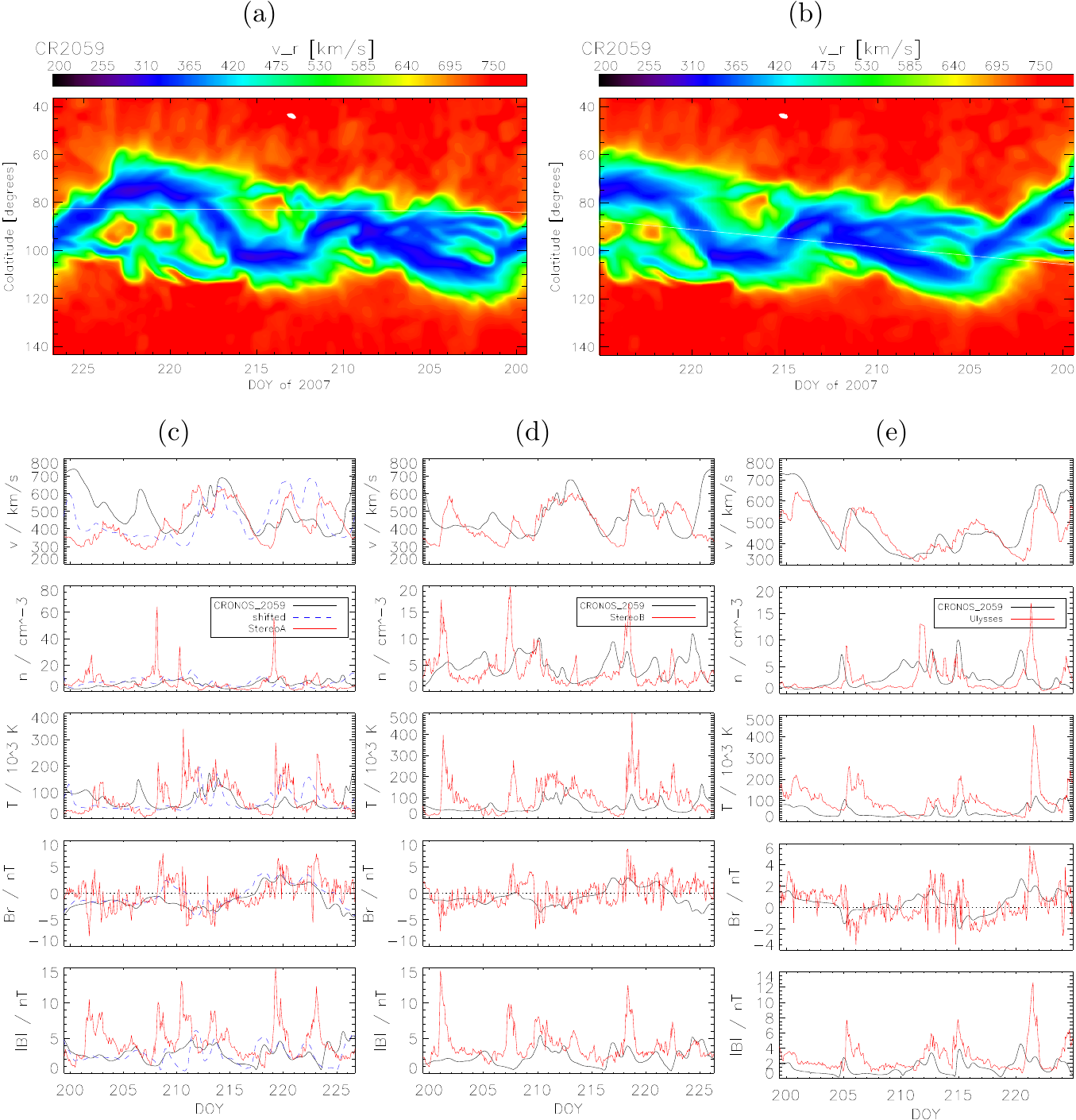}
\caption{Same as Figure \ref{fig:comp}, now for CR2059}
\label{fig:comp2}
\end{figure}
\begin{figure}
\plotone{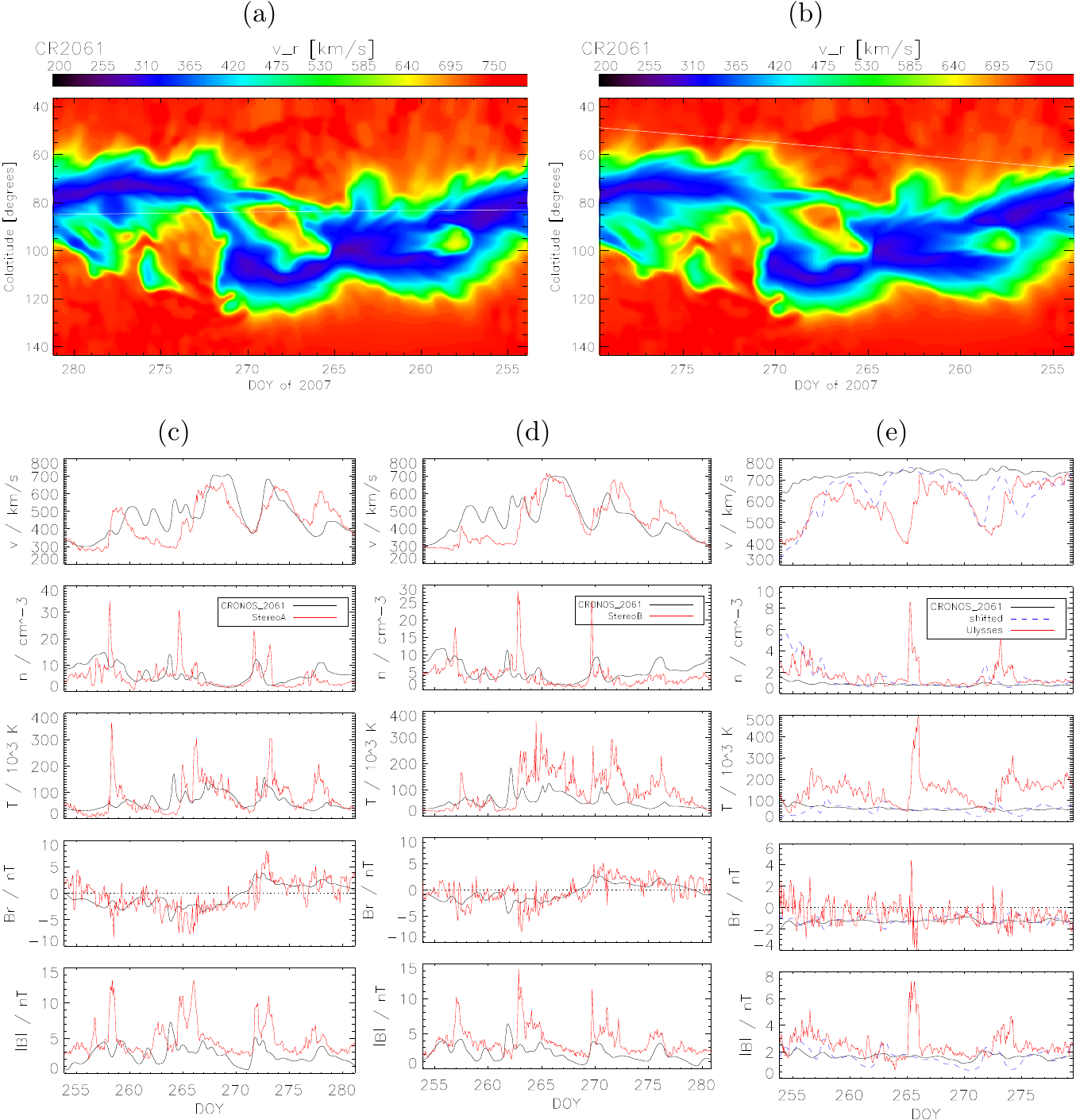}
\caption{Same as Figure \ref{fig:comp}, now for CR2061}
\label{fig:comp3}
\end{figure}
Figure \ref{fig:comp2} shows results for CR2059 in the same format as Figure \ref{fig:comp}. Most features are captured relatively well in the STEREO comparisons with the largest deviations occuring at the beginning and end of this CR where the simulation data show a high-speed feature not seen in the measurements. A shift in latitude for STEREO-A shows again that a small deviation from the actual trajectory gives better results and proves the presence of respective high-speed features, which are just slightly offset. The comparison with Ulysses data shows excellent agreement.\\
The results for CR2061 are shown in Figure \ref{fig:comp3} which are similar to the ones for CR2060 along the STEREO trajectories, and satisfactory agreement is found. Ulysses, on its way to northern polar regions, encounters predominantly the fast solar wind coming from the respective northern coronal hole. The simulation data along its trajectory, however, does not show the return to slow-speed wind occuring twice. Again, these are present in the global topology (see panel (b)) but are located too far south so that an 8$^\circ$ degree shift south in latitude is necessary to produce results more similar to the measurements.\\
We believe to have found a reasonable agreement with spacecraft data at relatively small heliocentric distances and, therefore, we feel that the results we obtain at greater distances can be trusted to also be a good representation to the local solar wind conditions. 
\begin{figure}
\begin{center}
\includegraphics[width=0.8\textwidth]{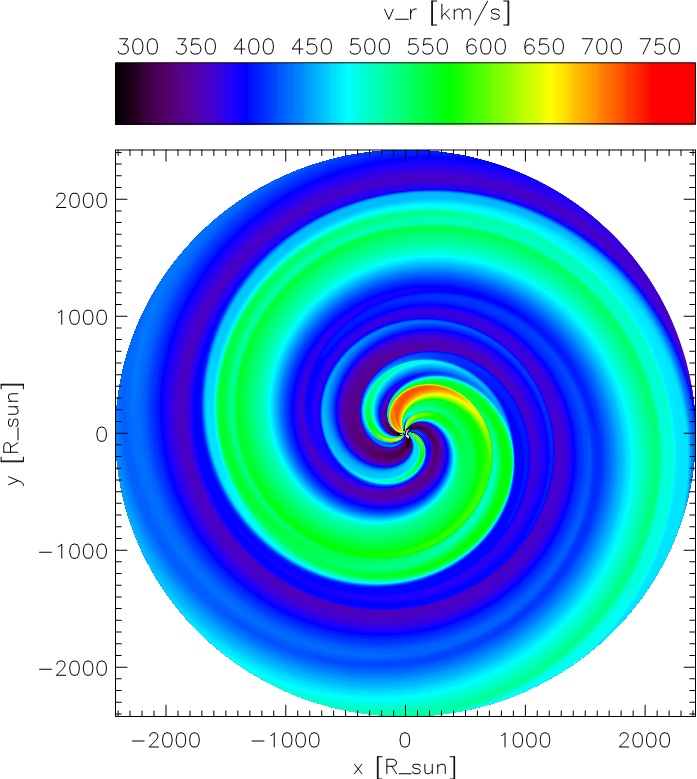}
\caption{Equatorial slice for CR2060 for a simulation extended to 10AU.} 
\label{fig:2060_eq}
\end{center}
\end{figure}
\begin{figure}
\begin{center}
\includegraphics[width=0.8\textwidth]{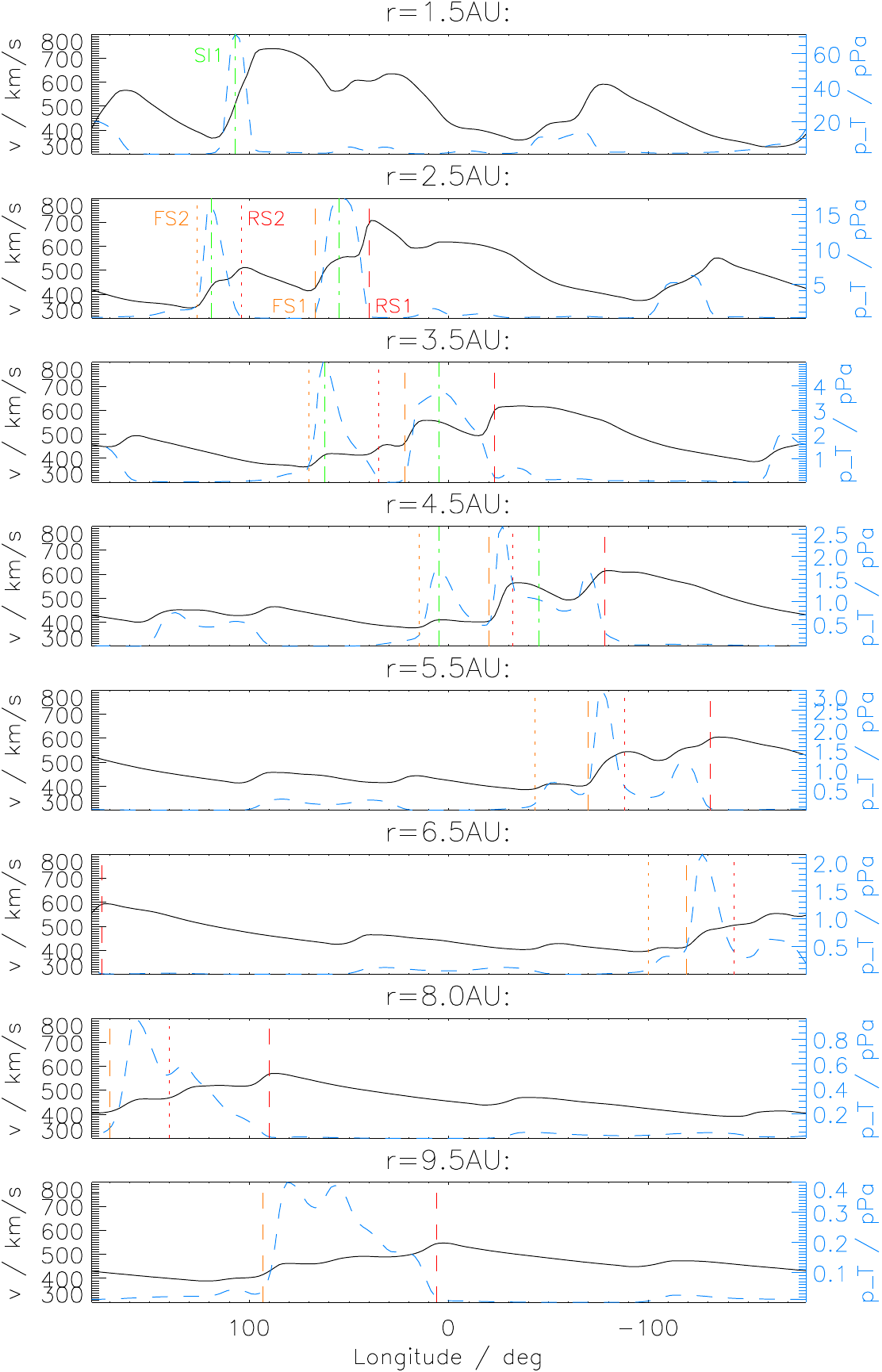}
\caption{Evolution and interaction of adjacent CIR structures in the equatorial plane from 1.5~AU to 9.5~AU for CR2060. The format is similar to Figure \ref{fig:pizzo_evolution}.} 
\label{fig:2060_evolution}
\end{center}
\end{figure}
An arising problem, however, is the common assumption of stationary boundary conditions during the course of a CR. On the one hand, this is reasonable for propagating solutions out to 1-2 AU only, because the solar wind takes a relatively short time to propagate there ($\approx4$ days/AU) as compared to the duration of a CR ($\approx27$ days). On the other hand, when extending the simulations to larger radial distances, stationary boundary conditions may become unreasonable if there is a significant change from one CR to the next one, because the propagation time becomes comparable to and eventually even larger than the duration of a CR, so that the heliosphere is filled with solar wind whose composition changes according to the different boundary conditions. This latter effect will have to be considered when we extend the simulation box to larger radial distances, so that time-dependent boundary conditions have to be applied. This will, however, take considerably longer computation time. A simpler approach appears to be possible for CR2060-2061, because the boundary conditions change rather little and stationary boundary conditions still appear to be a reasonable assumption. A respective simulation was carried out applying constant inner radial boundary conditions of CR2060 and extending the simulation box in the radial direction to 10AU with a resolution of $[\Delta r,\Delta\vartheta,\Delta\varphi]=[2R_\odot,2^\circ,2^\circ]$. \\
An equatorial slice of the simulation box depicting radial velocity is shown in Figure \ref{fig:2060_eq}. Three initially distinct high-speed streams can be identified that --- with increasing heliocentric distance --- begin to interact and form a CMIR with a large angular extent. This is shown in a quantitative manner in Figure \ref{fig:2060_evolution}, which is similar to Figure \ref{fig:pizzo_evolution} as it shows radial velocity (black solid) and total pressure (blue dashed) at the equator at radial distances from 1.5 to 9.5~AU. The horizontal axis uses a longitude, which is Carrington Longitude minus $180^\circ$ in order to show the evolution and interaction out to 6.5~AU without hitting the longitudinal periodic boundary. Only one of the three initial high-speed streams at 1.5~AU shows a strong compression region at its stream interface (SI1, green dashed-dotted). At 2.5~AU two stream interfaces with respective forward (FS orange, dashed/dotted) and reverse shocks (RS, red dashed/dotted) propagating away from them can be seen. FS1 and RS2 move towards and finally through each other between 3.5 and 4.5~AU and a CMIR is formed, which further interacts with the third initial high-speed stream's shocks a little beyond 8~AU. The only prominent structure left at 9.5~AU is a large merged interaction region bounded by the forward and reverse shocks (FS1 and RS1) of the initially steepest high-speed feature with a complicated internal structure as a result of the merging process. The influence of such complicated structures on particle transport will be interesting to investigate.

\section{Summary and Discussion}
\label{sec:conclusions}
We demonstrated the capability of the MHD code CRONOS to model CIR-associated structures (compression regions, shock pairs) in a test case where we are in agreement with the earlier results by \citet{Pizzo-1982}. To model more realistic solar wind conditions, we used GONG magnetograms and the FDIPS potential field solution as input to the WSA model to derive inner boundary conditions for our MHD code. To our knowledge this is the first time that the WSA model is used in conjuction with the FDIPS model, which can make use of the full resolution of a given magnetogram without introducing numerical artifacts that can arise in the usual spherical harmonics expansion of the coronal potential field. Our results could be shown to be in reasonable agreement with spacecraft data.\\
Other studies have looked at CR2060 using input from the WSA model. In one example, \citet{Pahud-etal-2012} validated their findings by comparison with ACE and MESSENGER spacecraft data, though no comparison with Ulysses data was performed to validate out-of-ecliptic results. The agreement of the in-ecliptic results is comparable to the one found here. Ulysses comparisons for this time period were performed by \citet{Broiles-etal-2013}, but the focus was on single CIR structures instead of investigating the whole CR. Therefore, in performing simulations to reproduce simultaneous multi-spacecraft observations including out-of-ecliptic data and also considering temporally adjacent CRs our modeling extends previous work and provides a suitable framework for a subsequent study of 3D energetic particle transport.\\ 
There are several reasons for occasional deviations when directly comparing simulation results to spacecraft in situ measurements which can essentially be attributed to the simplifications made in the model. Amongst others the following reasons can be listed: First, it has been shown that results using input from different solar observatories may differ quite substantially \citep{Pahud-etal-2012, Gressl-etal-2013}. Along this track \citet{Riley-etal-2013} investigated an ensemble modeling technique taking into account results from different models and observatories, which when combined give a better solution. Secondly, the empirical formulas used to set the inner radial boundary conditions are not well constrained and need some tuning that may depend on the observatories input data and the phase of the solar cycle. Similarly, the PFSS and SCS models are rather crude estimates of the inner and outer coronal fields, and are also subject to empirical parameters such as the source surface radius, which may not be constant as commonly assumed, but may vary depending on angular position \citep{Riley-etal-2006} and solar cycle \citep{Tran-2009}. A tuning of all these parameters could be performed to diminish differences between simulations and in-situ measurements, however, this does not give insight into the underlying physics on the one hand, and is also a very time consuming undertaking on the other hand. Thirdly, since the whole solar surface is not visible from Earth at a given time, synoptic magnetograms always contain non-simultaneous observations. Flux-transport models (e.g.~\citet{Jiang-etal-2010}) may be an effective tool to model the unobserved hemisphere of the Sun and improve magnetograms, however. These are not implemented in our model, but may be subject to future work.\\
A related problem is the common assumption of stationary boundary conditions during the course of a CR. While this remains reasonable for propagating solutions out to about 1~AU only --- because the solar wind's crossing time is much smaller than the duration of a CR --- it becomes necessary to apply time-dependent boundary conditions when extending the simulations to larger radial distances so that the heliosphere is filled with solar wind with composition changes according to the changing coronal conditions.\\
This latter effect will have to be considered when we extend the simulation box to larger radial distances, so that time-dependent boundary conditions may have to be applied. This will, however, take considerably longer computation time. A simpler approach presented here was taken for CR2060-2061, because the boundary conditions change rather little and stationary boundary conditions might still be a reasonable assumption. Time-dependent simulations and the inclusion of the poles will be adressed in an upcoming paper, where we will utilize the modeled 3D solar wind structure to investigate the transport of energetic particles, such as Jovian electrons and galactic cosmic rays.

\acknowledgments


Financial support for the project FI 706/8-2 (within Research Unit 1048), as well as for the projects FI 706/14-1 and HE 3279/15-1 funded by the Deutsche Forschungsgemeinschaft (DFG), and FWF-Projekt I1111
is acknowledged. The STEREO/SEPT Chandra/EPHIN and SOHO/EPHIN project is supported under grant 50OC1302 by the Federal Ministry of Economics and Technology on the basis of a decision by the German Bundestag. Tobias Wiengarten wants to thank Nick Arge and Janet Luhman for their helpful comments.



{\it Facilities:} \facility{NSO/GONG}.

\clearpage

\end{document}